\documentclass{jpsj3}

\usepackage{amsmath}
\usepackage{graphicx}

\def\nn{\nonumber}

\title{Layered Dynamical Conductivity for a Transfer Matrix Method \\
 $-$Application to an ${\cal N}$-layer Graphene$-$}

\author{Ken-ichi Sasaki}
\inst{NTT Research Center for Theoretical Quantum Physics and NTT Basic Research Laboratories, NTT Corporation,
3-1 Morinosato Wakamiya, Atsugi, Kanagawa 243-0198, Japan}

\abst{
 We calculated the optical properties of an $N$-layer graphene
 by formulating the dynamical conductivity of each layer.
 This is the conductivity when an electromagnetic field is localized at a particular layer 
 and differs from the standard conductivity calculated assuming a uniform field throughout all layers.
 By combining these conductivities with a transfer matrix method, 
 we took into account the spatial variation of the electromagnetic field caused by internal reflections.
 The results obtained from the two conductivities show that 
 similar peak structures originating from the interlayer electronic interaction appear
 in reflectance of an $N$-layer graphene at any $N$.
 The peak is inherent to the AB stacking and is not seen for the AA stacking, 
 and the peak corresponding to a sufficiently large $N$ is considered to the one observed for natural graphite.
 We also gave physical explanations of the existing experimental results 
 on highly oriented pyrolytic graphite and natural graphite under high pressure. 
 Although a layered conductivity underestimates the reflectance of graphite at photon energies below the peak, 
 we will show that 
 the disagreement is attributed to a nonlocal conductivity caused by interlayer interaction. 
 The calculations with layered conductivity are useful in knowing the local response to light and 
 may be further validated by an observation of a correction by interlayer electronic interaction to 
 the universal layer number that we have discovered recently.
 }

\begin{document}
\maketitle

\section{Introduction}

The exfoliation of a single-layer graphene from graphite 
provided a great opportunity to explore 
the behavior of massless Dirac fermions.~\cite{Novoselov2005,zhang2005}
The number of layers in graphite decreases by the mechanical cleavage method 
from a huge number to unity.~\cite{Novoselov2005a}
Considering the fact that graphite generally possesses massive Dirac fermions,
there should be interesting physics, relevant to a change in the ``mass'' of Dirac fermions
that is governed by the change in the layer number from $\infty$ to $1$.
The ``mass'' is closely related to the stacking order and interlayer distance 
which are changed by thermal expansion caused by absorption of light.
Therefore, it is meaningful to investigate the optical properties of an $N$-layer graphene 
as a function of $N$.
The potential of graphene expands with just one additional layer, 
especially when thermal instability is introduced.
This fact is evident from the appearance of many-body effects 
such as superconductivity and magnetism in a thermally unstable twisted bilayer
which are unseen in a single layer.~\cite{Cao2018,Cao2018a}

\begin{figure}[htbp]
 \begin{center}
  \includegraphics[scale=0.6]{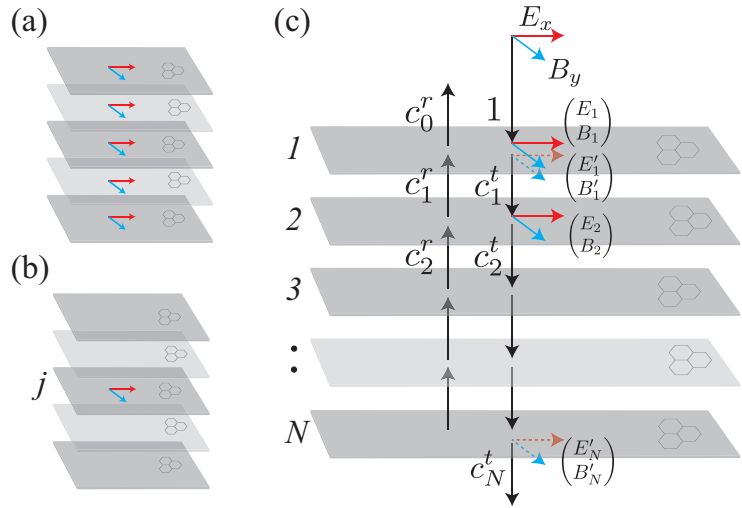}
 \end{center}
 \caption{(Color Online)
 (a) The bunched conductivity, $\sigma_N$, is defined 
 when the electromagnetic fields are sufficiently uniform in all layers.
 The electromagnetic fields are expressed by $E_x$ (red arrow) and $B_y$ (blue).
 (b) The layered conductivity of a $j$th layer in an $N$-layer graphene, 
 $\sigma^ N_j$, is calculated by assuming that the electric field exists only at the $j$th layer.
 (c) We apply $\sigma^ N_j$ to 
 the transfer matrix method for light propagation in an $N$-layer graphene.
 The interlayer distance $d$ is 0.335 nm.
 }
 \label{fig:sl}
\end{figure}

We have two different approaches to calculating the optical properties of an $N$-layer graphene
for light with normal incident on graphene plane.
One is a standard method, 
in which we calculate the dynamical conductivity, $\sigma_N$, 
by assuming that the electric field is spatially uniform in the layered material, 
as shown schematically in Fig.~\ref{fig:sl}(a).
Though the electronic states form standing wave in the $c$-axis direction due to interlayer coupling,
the current operator (in Kubo's formula) is invariant about the $c$-axis direction.
As a result, 
the optical matrix element is nonzero for only one or two specific final states against an initial state,
which is the selection rule of the wavenumber that makes the calculation simple.
Meanwhile, 
the adaptive range of this method using $\sigma_N$
is limited to the case that the resultant electromagnetic field is sufficiently uniform over the layers.
It is not straightforward to find a method to take into account 
the possible spatial variations of the electromagnetic field due to light absorption by $\sigma_N$ itself,
irregularities along the $c$-axis caused by defects, cracks and so on.

We may adopt a different approach 
in which the conductivity is calculated layer by layer, 
by assuming that an electric field localizes at only $j$th layer, as shown in Fig.~\ref{fig:sl}(b).
The dynamical conductivity 
$\sigma^N_j$ may vary depending on the layer position $j$ ($=1,\cdots,N$) 
because of interlayer electronic coupling.
Since the selection rule concerning the wavenumber along the $c$-axis is broken, 
the matrix element of the current operator of a layer includes all the possible final states 
that are energetically allowed.
As a result, the calculation becomes more complicated than $\sigma_N$.
However, the calculation based on $\sigma^N_j$ may have a wide application range.
For example, when light fields are enhanced in a specific layer
by plasmon resonance or by a metal layer as in the case of graphite intercalation compounds,
the fields around that layer are also enhanced.
Even in such a case, 
the optical characteristics of an $N$-layer graphene
can be estimated by combining 
$\sigma^N_{j=1,\cdots,N}$
with the transfer matrix method.

In this paper, 
we define the former $\sigma_N$ as a bunched conductivity
and the latter $\sigma^N_{j=1,\cdots,N}$ as a layered conductivity.
We quantified the differences between the two conductivities for the AB and AA stacking orders 
in the collisionless limit.
By comparing the calculated results from the two formulations for a sufficiently large $N$ value 
with reflectance measurements of graphite in a quantitative manner,
we point out a small deviation between theory and experiment.
The layered conductivity underestimates the reflectance of graphite at low photon energies.
It is explained by the effect of a nonlocal conductivity on the reflectance.
We then propose to investigate the absorptance as a function of $N$.
In a previous paper, 
we calculated the optical properties of an $N$-layer graphene and found that 
a characteristic peak structure appears in the absorption spectrum,
regardless of the photon energy, at the layer number $N=2/\pi \alpha \simeq 87$,
where $\alpha$ is the fine-structure constant.~\cite{Sasaki2020a}
This is a universal layer number found by assuming that any correlation between layers 
is negligible besides electromagnetic fields.
We will show that a correction by interlayer electronic interaction to the universal layer number
is sizable for the layered conductivity while it is modest for the bunched conductivity.
This feature may be used to test the validity of the layered conductivity.

This paper is composed as follows.
Basic knowledge about transfer matrix method and electronic states of an $N$-layer graphene 
with AB stacking order is provided in Sec.~\ref{sec:rev}.
We formulate those two dynamical conductivities in Sec.~\ref{sec:for}
and show the calculated results in Sec.~\ref{sec:cal}.
We conclude this paper in Sec.~\ref{sec:con}.
We examine an $N$-layer graphene with AB stacking order in the main test.
In Appendix~\ref{app:largeN},
we show an analytic expression for the bunched dynamical conductivity 
of an $\infty$-layer graphene (or graphite) with AB stacking order.
Some results for AA stacking are summarized in Appendix~\ref{app:aa}.

\section{Basic knowledge}\label{sec:rev}

\subsection{Transfer Matrix Method}\label{sec:tm}

We explain the transfer matrix method in Fig.~\ref{fig:sl}(c).~\cite{Mizuno1994}
The arrows along the $z$-axis indicate the propagation direction of light;
the first graphene layer transmits and reflects the incident light in the forward and backward directions
with certain amplitudes $c_1^t$ and $c_0^r$.
Such transmission and reflection are repeated at each layer. 
The absolute square of $c_N^t$ and $c_0^r$
corresponds to the transmittance $T_N$ and reflectance $R_N$, respectively.

We define electromagnetic fields
at an infinitesimal distance ($\epsilon$) above and below the $j$th layer as
\begin{align}
 \begin{pmatrix}
  E_{j} \cr B_{j}
 \end{pmatrix}
 \ {\rm and} \
 \begin{pmatrix}
  E'_{j} \cr B'_{j}
 \end{pmatrix},
\end{align}
respectively.
These are related by the boundary condition of the fields as
\begin{align}
 \begin{pmatrix}
  E'_{j} \cr B'_{j}
 \end{pmatrix}
 =
 \begin{pmatrix}
  1 & 0 \cr
  -\frac{\sigma^N_j}{\epsilon_0 c^2} & 1 
 \end{pmatrix}
 \begin{pmatrix}
  E_{j} \cr B_{j}
 \end{pmatrix}.
 \label{eq:bc}
\end{align}
The boundary condition is given by an integral of Maxwell's equations
(in differential form) over an infinitesimal interval 
$[z_j-\epsilon,z_j +\epsilon]$ containing the $j$th graphene.
The electric field is continuous ($E'_{j}=E_{j}$) according to Faraday's law, 
while the magnetic field is discontinuous ($B'_{j}=B_{j}-J^N_j/\epsilon_0 c^2$)
according to Amp\`ere's circuital law.
Since the discontinuity is given by the current of $j$th layer in an $N$-layer graphene
which is $J^N_j= \sigma^N_j E_j$, the layered dynamical conductivity $\sigma^N_j$
appears in the off-diagonal term of Eq.~(\ref{eq:bc}).

By assuming that electromagnetic fields with angular frequency of $\omega$ 
propagate freely in the interlayer vacuum space of distance $d$ by 
speed of light $c$, 
the electromagnetic fields between $j$ and $j+1$th layer ($z_{j} \le z \le z_{j+1}$)
[or at the $j$th interlayer space], 
are written in terms of the amplitudes $c^t_j$ and $c^r_j$ as 
$E_{j}(z)=c_j^t e^{i\omega z/c}+c_j^r e^{-i\omega z/c}$ 
and $cB_{j}(z)=c_j^t e^{i\omega z/c}-c_j^r e^{-i\omega z/c}$.
Therefore, electromagnetic fields at adjacent layers are related by
\begin{align}
 \begin{pmatrix}
  E_{j+1} \cr B_{j+1}
 \end{pmatrix}
 = 
\begin{pmatrix}
  \cos(\frac{\omega d}{c}) & i c \sin(\frac{\omega d}{c}) \cr
  \frac{i}{c} \sin(\frac{\omega d}{c})  & \cos(\frac{\omega d}{c})
 \end{pmatrix}
 \begin{pmatrix}
  E'_{j} \cr B'_{j}
 \end{pmatrix}.
 \label{eq:To}
\end{align}
We combine Eq.~(\ref{eq:bc}) and Eq.~(\ref{eq:To}) to define the transfer matrix
\begin{align}
 T_j
 = 
\begin{pmatrix}
  \cos(\frac{\omega d}{c}) & i c \sin(\frac{\omega d}{c}) \cr
  \frac{i}{c} \sin(\frac{\omega d}{c})  & \cos(\frac{\omega d}{c})
 \end{pmatrix}
 \begin{pmatrix}
  1 & 0 \cr
  -\frac{\sigma^N_j}{\epsilon_0 c^2} & 1 
 \end{pmatrix}
\label{eq:tmatrix}
\end{align}
that satisfies
\begin{align}
 \begin{pmatrix}
  E_{j+1} \cr B_{j+1}
 \end{pmatrix}
 =T_j
 \begin{pmatrix}
  E_j \cr B_j
 \end{pmatrix}.
 \label{eq:tmatr}
\end{align}
The matrix is obtained from Maxwell's equations and is the product of two matrices. 
The first matrix expresses the propagation of light in the interlayer space and
the second matrix represents the boundary condition at a graphene layer.
Our approaches are based on the approximation 
that the electronic current is sufficiently localizing at a graphene layer.
But in fact,
because electronic wave function of the $\pi$-orbital is slightly spreading in a space between graphene layers,
light propagation in the space is likely to be subjected to the spread of the wave function.
The possible effects outside of this approximation may be examined
by introducing an appropriate dielectric constant of the space.

By multiplying the transfer matrix with the field at an infinitesimal distance above the top layer
$N-1$ times, we obtain the field at an infinitesimal distance below the $N$th layer,
\begin{align}
 \begin{pmatrix}
  E'_{N}(z_N) \cr B'_{N}(z_N)
 \end{pmatrix}
 =
 \begin{pmatrix}
  1 & 0 \cr
  -\frac{\sigma^N_N}{\epsilon_0 c^2} & 1 
 \end{pmatrix}
 T_{N-1}\cdots T_2 T_1
 \begin{pmatrix}
  E_{1}(z_1) \cr B_{1}(z_1)
 \end{pmatrix}.
 \label{eq:ebmatrix}
\end{align}
We normalize the amplitude of the incident light to unity ($c_0^t=1$);
$E_{1}(z_1)=e^{i\omega z_1/c}+c_0^r e^{-i\omega z_1/c}$ and 
$cB_{1}(z_1)=e^{i\omega z_1/c}-c_0^r e^{-i\omega z_1/c}$ holds for the light at the entrance.
Because $c_N^r$ should vanish for the field at the exit,
$E'_{N}(z_N)=c_N^t e^{i\omega z_N/c}$ and $cB'_{N}(z_N)=c_N^t e^{i\omega z_N/c}$ 
or $E'_{N}(z_N)=cB'_{N}(z_N)$ must hold.
Thus, Eq.~(\ref{eq:ebmatrix}) provides two equations 
for determining the amplitudes $c_0^r$ and $c_N^t$, by which we obtain $R_N$ and $T_N$.

The total absorptance is given by $A^N=1-R_N-T_N$.
By defining a Pointing vector, 
it is straightforward to show using Eq.~(\ref{eq:bc}) 
that $\epsilon_0 c^2 (E'_j)^* B'_j=\epsilon_0 c^2 E_j^* B_j - \sigma_j^N |E_j|^2$ 
(where $E_j^*$ is the complex conjugate of $E_j$)
and that $A^N$ is equivalent to the sum of the energies absorbed by each graphene layer 
(layer absorptance $A^N_j\equiv \sigma^N_j |E_j|^2$),
\begin{align}
 A^N = \sum_{j=1}^N \sigma^N_j |E_j|^2.
 \label{eq:AN}
\end{align}
The field configuration $E_j$ is locally determined 
($j=1,\cdots,N$) by a given set of $\sigma^N_{j=1,\cdots,N}$
because once $c_0^r$ is known, $c^t_i$ and $c^r_i$ can be calculated
by using Eq.~(\ref{eq:tmatr}) repeatedly.
Therefore, $A^N$ is actually a nonlinear equation of $\sigma^N_j$.

The transfer matrix method is applicable to the bunched conductivity.~\cite{Heavens1960}
Because light propagates ``freely'' through the medium with 
the relative permittivity 
\begin{align}
 \varepsilon_N(\omega) \equiv 1 + i \frac{\sigma_N(\omega)}{\epsilon_0 \omega},
\end{align}
we have 
\begin{align}
 \begin{pmatrix}
  E'_{N}(z_N) \cr B'_{N}(z_N)
 \end{pmatrix}
 =
\begin{pmatrix}
  \cos(\sqrt{\varepsilon_N} \frac{\omega (z_N-z_1)}{c}) & i \frac{c}{\sqrt{\varepsilon_N}} \sin(\sqrt{\varepsilon_N}\frac{\omega (z_N-z_1)}{c}) \cr
  \frac{i\sqrt{\varepsilon_N}}{c} \sin(\sqrt{\varepsilon_N}\frac{\omega (z_N-z_1)}{c})  & \cos(\sqrt{\varepsilon_N}\frac{\omega (z_N-z_1)}{c})
 \end{pmatrix}
 \begin{pmatrix}
  E_{1}(z_1) \cr B_{1}(z_1)
 \end{pmatrix},
\end{align}
instead of Eq.~(\ref{eq:ebmatrix}). 
The $2\times 2$ matrix is given in Eq.~(\ref{eq:To}) 
by replacing $d$ with $z_N-z_1$ and $c$ with $c/\sqrt{\varepsilon_N}$.
We assume that the electromagnetic fields are continuous at the interface 
between the air and the material.
By eliminating $c^t_N$ from the above equation, we obtain
\begin{align}
 c^r_0 = \frac{i\left(\sqrt{\varepsilon_N}- \frac{1}{\sqrt{\varepsilon_N}} \right)\sin\left(\sqrt{\varepsilon_N} \frac{\omega (z_N-z_1)}{c}\right)}{2\cos\left(\sqrt{\varepsilon_N} \frac{\omega (z_N-z_1)}{c}\right) - i \left(\sqrt{\varepsilon_N}+ \frac{1}{\sqrt{\varepsilon_N}} \right)\sin\left(\sqrt{\varepsilon_N} \frac{\omega (z_N-z_1)}{c}\right)}.
 \label{eq:cr0}
\end{align}
From which we can calculate reflectance and transmittance.
When $N$ is sufficiently large, 
this result reproduces the standard formula, $R=|(\sqrt{\varepsilon_N}-1)/(\sqrt{\varepsilon_N}+1)|^2$ 
or $R=[(n-1)^2+\kappa^2]/[(n+1)^2+\kappa^2]$, 
where optical constants were determined by $\varepsilon_N = (n+i\kappa)^2$,
namely, 
\begin{align}
 n(\omega)= \sqrt{\frac{1+\sqrt{1+ \left( \frac{\sigma_N(\omega) c}{\omega} \right)^2 }}{2} }, \ \
 \kappa(\omega)= \frac{\sigma_N(\omega) c}{2n(\omega) \omega}.
\end{align}

\subsection{Electronic Properties}

We review the electronic properties of an $N$-layer graphene with AB stacking order.~\cite{Min2008}
Due to the reflections of electrons taking place at the first and $N$th surface layers 
caused by the broken translation symmetry of the lattice, 
the electron standing wave is formed and the state 
is characterized by the wavenumber along the $c$-axis as,
\begin{align}
 k_r = \frac{r\pi}{N+1}, \  \ (r =1,\cdots,N).
\end{align}
For each layer,
we adopt the model of massless Dirac fermions with a linear energy dispersion $\pm vp$,
where $p$ is the magnitude of two dimensional (in plane or lateral) wavevector ${\bf p}$
and $v$ the Fermi velocity.
Hereafter we use a unit in which $v$ is unity,
because the calculated results are independent of it.
Note that we neglect lateral standing waves within a layer 
that appear near the edge of a graphene layer 
by assuming that each layer is sufficiently large.~\cite{Sasaki2011}

The effect of the hopping integral $\gamma_1$ between the nearest layers
on massless Dirac fermions
can be taken into account as a ``mass'' of Dirac fermions,
\begin{align}
 m_r = \gamma_1 \cos(k_r).
 \label{eq:mass}
\end{align}
The energy dispersion relation of the massive Dirac fermions becomes
\begin{align} 
 \varepsilon_{r{\bf p}}^{s} = m_r + s \sqrt{p^2 + m_r^2},
\end{align}
where the positive and negative energy eigenstates are separated by the band index $s=\pm$;
i.e., $\varepsilon_{r{\bf p}}^{+}\ge 0$ and $\varepsilon_{r{\bf p}}^{-}\le 0$.
The mass appears as a (normal) mass that creates an energy bandgap $s\sqrt{p^2 + m_r^2}$
and also as a potential that shifts the band center by $m_r$.
The ``mass'' shift can be positive and negative depending on the $k_r$ value.
We assume a positive value of $\gamma_1$ in this paper, though the negative one is more plausible.
No result is changed by this convention because the difference is removed by the replacement 
$k_r \to k_r + \pi$.

Dirac fermions acquire different kinds of masses, 
depending on the patterns of the symmetry breaking of the equivalence between two carbon atoms 
in the hexagonal unit cell (A and B atoms, known as pseudospin).
By combining the concept of the ``mass'' with valley and spin degrees of freedoms, 
we can argue in a unified manner various aspects of physics from intriguing phenomena 
such as the quantum Hall effect~\cite{haldane88} and quantum spin Hall effect~\cite{Kane2005,Fu2007}
to the bandgap engineering.~\cite{Novoselov2007} 
As we will show later, 
the mass of Eq.~(\ref{eq:mass}) is indeed 
the most important quantity that governs the dynamical conductivity 
for photon energy of interest.

The wavefunction can be read from the following basic pattern of the first and second layers,
\begin{align}
 |\Psi_{r{\bf p}}^s \rangle 
 =
 N_{r{\bf p}}^s
 \begin{pmatrix}
  \sin(k_r) & 0 & 0 & 0 & \cdots \cr
  0 & \sin(k_r) & 0 & 0 & \cdots \cr
  0 & 0 & \sin(2k_r) & 0 & \cdots \cr
  0 & 0 & 0 & \sin(2k_r) & \cdots \cr
  \vdots & \vdots & \vdots & \vdots & 
 \end{pmatrix}
 \begin{pmatrix}
  \frac{\Pi^*}{\varepsilon_{r{\bf p}}^s} \cr
  1 \cr
  1 \cr
  \frac{\Pi}{\varepsilon_{r{\bf p}}^s} \cr
  \vdots
 \end{pmatrix},
 \label{eq:wf}
\end{align}
where $\Pi \equiv p_x + i p_y$ and $\Pi^* \equiv p_x - i p_y$.
The probability amplitude of the $j$th layer having two components structure (pseudospin)
is proportional to $\sin(jk_r)$.
The diagonal components of the $2N \times 2N$ matrix show that 
the wavefunction is the symmetrical state for the case of a positive mass $m_r > 0$,
while it is the antisymmetrical state for a negative mass $m_r <0$.
Since degenerate symmetrical and antisymmetrical states 
can be superimposed to make the wavefunction vanish at every two layers,
there is a state that does not feel the interlayer interaction.
This degeneracy is possible when $N$ is an odd number, and 
such a state with the original massless linear dispersion exists.

Because $\sum_{j=1}^N \sin^2(j k_r) = \frac{N+1}{2}$ holds,
the normalization constant $N_{r{\bf p}}^s$ in Eq.~(\ref{eq:wf}) is given by
\begin{align}
 (N_{r{\bf p}}^{s})^2 
 = \frac{(\varepsilon_{r{\bf p}}^{s})^2}{(\varepsilon_{r{\bf p}}^{s})^2+p^2} \frac{2}{N+1} 
 = \frac{\varepsilon_{r{\bf p}}^{s}}{sE_r} \frac{1}{N+1}.
\end{align}
In the last equation, we have introduced the energy of a (normal) massive Dirac fermions 
$E_r \equiv \sqrt{p^2 + m_r^2}$,
which will be used frequently in the subsequent calculations.

\section{Conductivity calculation}\label{sec:for}

\subsection{Layered Conductivity}

We define the layered dynamical conductivity of the $j$th layer in an $N$-layer graphene,
within a linear response theory, as
\begin{align}
 \sigma_j^N(\omega) = g_s g_v \frac{\hbar}{iS} \sum_{s,s'} \sum_{\bf p} \sum_{r,r'}
 \frac{f(\varepsilon_{r'{\bf p}}^{s'})-f(\varepsilon_{r{\bf p}}^{s})}{(\varepsilon_{r'{\bf p}}^{s'}-\varepsilon_{r{\bf p}}^{s})(\varepsilon_{r'{\bf p}}^{s'}-\varepsilon_{r{\bf p}}^{s}+\hbar \omega + i\epsilon)}
|\langle \Psi_{r{\bf p}}^{s}| (-e\hat{v}_j) |\Psi_{r'{\bf p}}^{s'}\rangle|^2.
 \label{eq:layeredsigma}
\end{align}
Here, $g_s(=2)$ and $g_v(=2)$ represents spin and valley degeneracy, respectively,
and $S$ the area of the graphene layer.
We take the limit $S\to \infty$ in the following calculations.
Multiplying this conductivity with the electric field at the $j$th layer 
gives the local current in an $N$-layer graphene $J^N_j = \sigma_j^N E_j$. 
Therefore, the layered conductivity fits the framework of the transfer matrix method.

Because the velocity operator $\hat{v}_j$
is written in terms of Pauli matrices $\sigma_x$ or $\sigma_y$ for pseudospin 
depending on the light polarization direction, the current matrix element becomes
\begin{align}
 |\langle \Psi_{r{\bf p}}^{s}|(-e \hat{v}^\pm_j) |\Psi_{r'{\bf p}}^{s'}\rangle|^2
 =e^2 (N_{r'{\bf p}}^{s'})^2 (N_{r{\bf p}}^{s})^2 
 \sin^2(jk_r) \sin^2(jk_{r'}) 
 \left|\frac{\Pi}{\varepsilon_{r'{\bf p}}^{s'}} \pm \frac{\Pi^*}{\varepsilon_{r{\bf p}}^{s}} \right|^2,
\end{align}
where the $\pm$ signs denote $x$ and $y$ polarization.
The dependence of the matrix element on the polarization is eventually lost 
for the dynamical conductivity, by the integral about polar angle of ${\bf p}$.
Putting above into Eq.~(\ref{eq:layeredsigma}) gives
\begin{align}
 \sigma^N_j(\omega) &= g_s g_v \frac{e^2}{i\hbar} \sum_{s,s'} \frac{1}{2\pi} \int pdp \sum_{r,r'}
 \frac{f(\varepsilon_{r'{\bf p}}^{s'})-f(\varepsilon_{r{\bf p}}^{s})}{(\varepsilon_{r'{\bf p}}^{s'}-\varepsilon_{r{\bf p}}^{s})(\varepsilon_{r'{\bf p}}^{s'}-\varepsilon_{r{\bf p}}^{s}+\hbar \omega + i\epsilon)} \nn \\
 & \times \frac{1}{sE_{r}} \frac{\sin^2(jk_r)}{N+1} \times
 \frac{1}{s'E_{r'}} \frac{\sin^2(jk_{r'})}{N+1} \times
 p^2 \left(\frac{\varepsilon_{r{\bf p}}^{s}}{\varepsilon_{r'{\bf p}}^{s'}} + \frac{\varepsilon_{r'{\bf p}}^{s'}}{\varepsilon_{r{\bf p}}^{s}} \right).
\end{align}
We evaluate this for charge neutral condition $E_F=0$ and at zero temperature $T=0$.
Then, the Fermi distribution functions become the step functions. 
As a result, since $\varepsilon_{r{\bf p}}^{+} \ge 0$ and $\varepsilon_{r{\bf p}}^{-} \le 0$,
only interband transitions contribute to the dynamical conductivity;
\begin{align}
 \sigma^N_j(\omega)
 &= -g_s g_v \frac{e^2}{i\hbar} \frac{1}{2\pi} \int pdp \sum_{r,r'}
 \frac{1}{(\varepsilon_{r'{\bf p}}^{-}-\varepsilon_{r{\bf p}}^{+})(\varepsilon_{r'{\bf p}}^{-}-\varepsilon_{r{\bf p}}^{+}+\hbar \omega + i\epsilon)} \nn \\
 & \times \frac{1}{E_{r}} \frac{\sin^2(jk_r)}{N+1} \times
 \frac{1}{E_{r'}} \frac{\sin^2(jk_{r'})}{N+1} \times
 p^2 \left(\frac{\varepsilon_{r{\bf p}}^{+}}{\varepsilon_{r'{\bf p}}^{-}} + \frac{\varepsilon_{r'{\bf p}}^{-}}{\varepsilon_{r{\bf p}}^{+}} \right).
\end{align}
By taking the collisionless limit ($\epsilon \to 0$), we use
$\left( \varepsilon_{r'{\bf p}}^{-}-\varepsilon_{r{\bf p}}^{+}+\hbar \omega + i\epsilon \right)^{-1} = -i\pi 
\delta \left( \varepsilon_{r'{\bf p}}^{-}-\varepsilon_{r{\bf p}}^{+}+\hbar \omega \right)$.
Moreover, because the Dirac delta function may be rewritten as
\begin{align}
 \delta \left( \frac{\varepsilon_{r'{\bf p}}^{-}-\varepsilon_{r{\bf p}}^{+}}{\partial p}|_{p=p_{rr'}} (p-p_{rr'}) \right),
\end{align}
where $p_{rr'}$ satisfies $\varepsilon_{r'{\bf p}_{rr'}}^{-}-\varepsilon_{r{\bf p}_{rr'}}^{+}+\hbar \omega=0$,
we obtain after integrating over $p$, 
\begin{align}
 \sigma^N_j(\omega) &= - g_s g_v\frac{e^2}{4\hbar}
 \sum_{r,r'} \frac{p_{rr'}^2}{\hbar \omega |E^2_{r'}(r) + E^1_{r}(r')|}
 \frac{\sin^2(jk_r)}{N+1}
 \frac{\sin^2(jk_{r'})}{N+1} \nn \\
 &\times
 \left( \frac{m_r + E^1_{r}(r')}{m_{r'} - E^2_{r'}(r)} + \frac{m_{r'} - E^2_{r'}(r)}{m_r + E^1_{r}(r')} \right)
 \Theta(E^1_{r}(r')-m_r)\Theta(E^2_{r'}(r)+m_{r'}),
 \label{eq:stepsigma}
\end{align}
where
\begin{align}
 & 
 E^1_{r}(r') \equiv E_r(p_{rr'}) = \frac{(\hbar \omega + m_{r'} - m_r )^2 - (m_{r'}-m_r)(m_{r'}+m_r)}{2(\hbar \omega + m_{r'} - m_r )},
 \\
 &
 E^2_{r'}(r) \equiv E_{r'}(p_{rr'}) = \frac{(\hbar \omega + m_{r'} - m_r )^2 + (m_{r'}-m_r)(m_{r'}+m_r)}{2(\hbar \omega + m_{r'} - m_r )}.
\end{align}
The appearance of the step functions in Eq.~(\ref{eq:stepsigma}), 
satisfying $\Theta(x) = 1$ for $x \ge 0$ and 0 otherwise, needs explanations.
They represent the conditions of the existence of $p_{rr'}$.
Since $\varepsilon_{r{\bf p}}^{+} \ge 0$ for the conduction band, 
$m_r + E^1_r(p_{rr'}) \ge 0$ holds.
On the other hand, since $p_{rr'}^2 = E^1_{r}(r')^2 -m_r^2$,
$E^1_{r}(r') - m_r \ge 0$ must be satisfied in order that $p_{rr'}$ exists.
Similarly,
since $\varepsilon_{r'{\bf p}}^{-} \le 0$ for the valence band, 
$m_{r'} - E^2_{r'}(p_{rr'}) \le 0$ holds.
On the other hand, since $p_{rr'}^2 = E^2_{r'}(r)^2 - m_{r'}^2$,
$E^2_{r'}(r) + m_{r'} \ge 0$ must be satisfied in order that $p_{rr'}$ exists.

Finally, we get a formula of the layered conductivity for $j=1,\cdots,N$,
\begin{align}
 \sigma^N_j(\omega) = g_s g_v \frac{e^2}{\hbar}
 \sum_{r,r'} \frac{m_r m_{r'} + E^1_{r}(r')E^2_{r'}(r)}{\hbar \omega |E^2_{r'}(r) + E^1_{r}(r')|}
 \frac{\sin^2(jk_r)}{N+1}
 \frac{\sin^2(jk_{r'})}{N+1}
 \Theta(E^1_{r}(r')-m_r)\Theta(E^2_{r'}(r)+m_{r'}).
 \label{eq:sigma_j}
\end{align}
For the general $N$, we evaluate Eq.~(\ref{eq:sigma_j}) numerically.
For a small $N$, we can evaluate Eq.~(\ref{eq:sigma_j}) analytically.
Particularly for the monolayer $N=1$, this formula reproduces the result
\begin{align}
 \sigma^1_1 = g_s g_v \frac{e^2}{16\hbar}=\pi \alpha \epsilon_0 c,
\end{align}
which is $\omega$-independent.
For the bilayer $N=2$, $\sigma^2_1(\omega)$ and $\sigma^2_2(\omega)$ are the same 
because $\sigma^N_{j}(\omega) = \sigma^N_{N+1-j}(\omega)$ holds, and 
\begin{align}
 & \sigma^2_1(\omega) = \sigma^2_2(\omega) = g_s g_v \frac{e^2}{16\hbar} \times \nn \\
 & \frac{1}{4}\left\{ 
 \left[
 1 + \frac{\gamma_1}{\hbar \omega+\gamma_1} \right] + 
 2 \left[
 1+ \left( \frac{\gamma_1}{\hbar \omega}\right)^2 \right]
 \Theta(\hbar \omega-\gamma_1)
 + \left[
 1- \frac{\gamma_1}{\hbar \omega-\gamma_1} \right]
 \Theta(\hbar \omega-2\gamma_1)
 \right\}.
\end{align}
We note that $\sigma^2_1(0)$ is exactly the half of $\pi \alpha \epsilon_0 c$.
The cases $N=1$ and $2$ are shown in Fig.~\ref{fig:sigma}(left panel).
When $\gamma_1=0$, the mass vanishes and 
$\sigma^N_{j=1,\cdots,N} = \pi \alpha \epsilon_0 c$ for arbitrary $N$.

\subsection{Bunched Conductivity}

When the electromagnetic fields are assumed to be sufficiently uniform in all layers,
the dynamical conductivity of an $N$-layer graphene is well approximated by 
\begin{align}
 \sigma_{N}(\omega) = g_s g_v
 \frac{\hbar}{iV} \sum_{s,s'} \sum_{\bf p} \sum_{r,r'}
 \frac{f(\varepsilon_{r'{\bf p}}^{s'})-f(\varepsilon_{r{\bf p}}^{s})}{(\varepsilon_{r'{\bf p}}^{s'}-\varepsilon_{r{\bf p}}^{s})(\varepsilon_{r'{\bf p}}^{s'}-\varepsilon_{r{\bf p}}^{s}+\hbar \omega + i\epsilon)} 
 |\langle \Psi^{s}_{r{\bf p}} | (-e\hat{v}) | \Psi^{s'}_{r'{\bf p}} \rangle|^2.
\end{align}
As regards this bunched conductivity, there is no concept of the local conductivity at a layer, 
as opposed to the layered conductivity.
Rather, we treat an $N$-layer graphene as a whole.

In the following, we consider $x$-polarization only, 
because it can be shown that 
there is no polarization dependence of the dynamical conductivity.
Using Eq.~(\ref{eq:wf}) and the following equalities
\begin{align}
 & \sum_{j=1}^N \sin(jk_{r'}) \sin(jk_r) = \frac{N+1}{2} \delta_{r',r}, \\
 & \sum_{j=1}^N (-1)^j \sin(jk_{r'}) \sin(jk_r) = - \frac{N+1}{2} \delta_{r',N+1-r},
\end{align}
we obtain absolute square of the matrix element of the current operator as
\begin{align}
 |\langle \Psi^{s}_{r{\bf p}} | (-e\hat{v}^+) | \Psi^{s'}_{r'{\bf p}} \rangle|^2 = e^2
 \frac{\varepsilon_{r'{\bf p}}^{s'}}{2s'E_{r'}} \frac{\varepsilon_{r{\bf p}}^{s}}{2sE_r}
 \left|{\rm Re} \left( 
 \frac{\Pi}{\varepsilon_{r'{\bf p}}^{s'}} + \frac{\Pi^*}{\varepsilon_{r{\bf p}}^{s}} \right) \delta_{r',r}
 + i {\rm Im} \left( 
 \frac{\Pi}{\varepsilon_{r'{\bf p}}^{s'}} + \frac{\Pi^*}{\varepsilon_{r{\bf p}}^{s}} \right)
 \delta_{r',N+1-r}
 \right|^2.
 \label{eq:mat2}
\end{align}
We focus on the interband transitions ($s=+1$ and $s'=-1$) 
with the assumption that $E_F=0$ and $T=0$.
Two different interband transitions contribute to $\sigma_{N}(\omega)$.
One originates from ${\rm Re}(\cdots)$ in Eq.~(\ref{eq:mat2}) which is associated 
with the direct transitions that preserve the wavenumber $(r'=r)$.
The other originates from ${\rm Im}(\cdots)$ which is associated 
with the indirect transitions that satisfy $r'=N+1-r$ (i.e., $k_r'+k_r=\pi$)
and the indirect interband transitions are associated with 
the change in the sign of the mass;
$m_{N+1-r} = - m_r$.
[When $N$ is an odd number, 
the indirect transitions contain a direct transition as a special case of $r'=r$, for which 
the mass vanishes.]
The former gives
\begin{align}
 \sigma_{N}^{\rm direct}(\omega) 
 &=g_s g_v
 \frac{e^2\hbar}{V} \sum_{\bf p} \sum_{r}
 \frac{\pi}{2E_r}
 \delta (\hbar \omega - 2 E_r)
 \frac{\varepsilon_{r{\bf p}}^{-}}{2E_r}
 \frac{\varepsilon_{r{\bf p}}^{+}}{2E_r}
 (p_x)^2 \left( 
 \frac{1}{m_r -E_r} + \frac{1}{m_r + E_r} \right)^2 \nn \\
 &=g_s g_v
 \frac{S}{V}\frac{e^2}{16\hbar} \sum_{r}
 \left( \frac{2m_r}{\hbar \omega} \right)^2 \Theta (\hbar \omega - 2 |m_r|).
 \label{eq:direct}
\end{align}
The step function represents the fact that direct
interband transitions exist for the limited photon energy above the bandgap, 
$\hbar \omega > 2 |m_r|$.
Meanwhile the latter gives
\begin{align}
 \sigma_{N}^{\rm indirect}(\omega) 
 &=g_s g_v
 \frac{e^2\hbar}{V} \sum_{\bf p} \sum_{r}
 \frac{\pi \delta (\hbar \omega - 2 (m_r +E_r))}{2(m_r +E_r)} 
 \frac{\varepsilon_{N+1-r{\bf p}}^{-}}{2E_r}
 \frac{\varepsilon_{r{\bf p}}^{+}}{2E_r} 
 (p_y)^2 \left( 
 \frac{1}{-m_r -E_r} - \frac{1}{m_r + E_r} \right)^2 \nn \\
 &=g_s g_v
 \frac{S}{V}
 \frac{e^2}{16\hbar} \sum_{r}\left(1- \frac{2m_r}{\hbar \omega -2m_r}\right)\Theta (\hbar \omega - 2 (m_r+|m_r|)).
 \label{eq:indirect}
\end{align}
It looks as if $\sigma_{N}^{\rm indirect}(\omega)$ can be singular at $\hbar \omega = 2m_r$.
However, due to the step function, $\sigma_{N}^{\rm indirect}(\omega)$ 
is a sum of (at most) discontinuous functions.

By combining Eqs.~(\ref{eq:direct}) and (\ref{eq:indirect}), 
we obtain a simple formula of the bunched conductivity as
\begin{align}
 \sigma_{N}(\omega) = \frac{S}{V} g_s g_v
 \frac{e^2}{16\hbar} \sum_{r=1}^N \left\{
 \left(1- \frac{2m_r}{\hbar \omega -2m_r}\right)\Theta (\hbar \omega - 2 (m_r+|m_r|))
 + \left( \frac{2m_r}{\hbar \omega} \right)^2 \Theta (\hbar \omega - 2 |m_r|) \right\}.
 \label{eq:bunchedc}
\end{align}
Here, $V$ denotes the three dimensional volume of the system,
$V=S$ for $N=1$ and $V=N S d$ for $N\ge 2$.
This formula reproduces the results obtained previously for $N=2$~\cite{Abergel2007,Koshino2008}
and for small $N \le 10$.~\cite{Min2009,Orlita2010}
The case of (AB stacking) graphite, 
i.e., $\lim_{N\to \infty} \sigma_N(\omega)$ can be analytically calculated and 
the result is shown in Appendix~\ref{app:largeN}.
Our result differs slightly from the calculations obtained previously by several authors.~\cite{Boyle1958,Ichikawa1966}

\section{Results and Discussion}\label{sec:cal}

\begin{figure}[htbp]
 \begin{center}
  \includegraphics[scale=1.0]{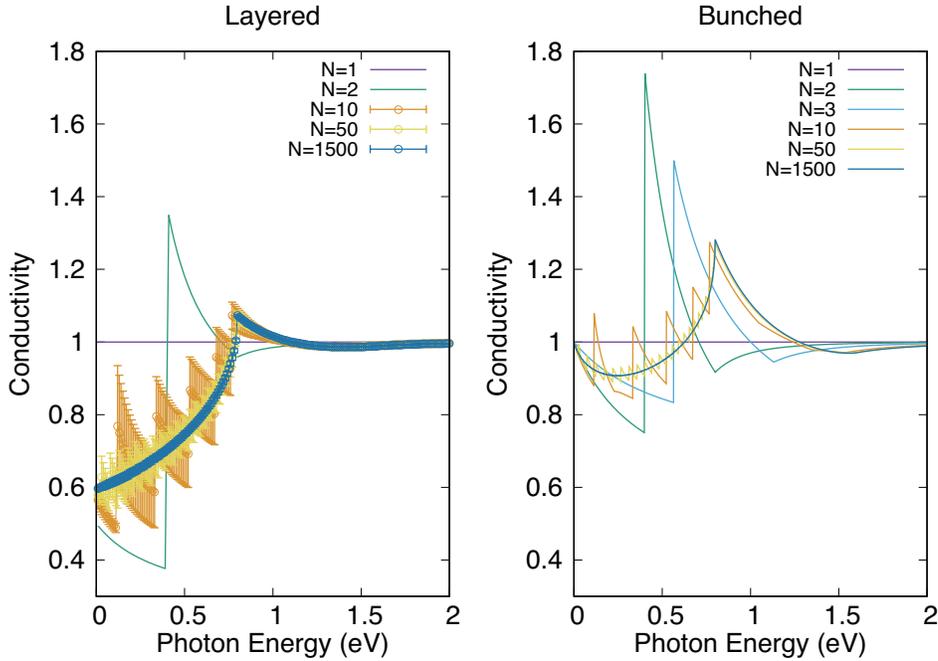}
 \end{center}
 \caption{(Color Online) Layered (left) and bunched (right) conductivities of an $N$-layer graphene
 as a function of photon energy $\hbar \omega$.
 The layered and bunched conductivities are scaled by $\sigma^1_1$ or $\sigma_1 d$, respectively.
 The error bars denote the standard deviation.
 }
 \label{fig:sigma}
\end{figure}

We plot the calculated conductivities as a function of photon energy
in Fig.~\ref{fig:sigma}.
In the left panel,
the layered conductivity is represented by 
the mean conductivity $\langle \sigma^N(\omega) \rangle \equiv \sum_{j=1}^N \sigma^N_j(\omega)/N$ and 
the standard deviation $\sqrt{\frac{1}{N} \sum_{j=1}^N (\sigma^N_j(\omega)-\langle \sigma^N(\omega) \rangle)^2}$.
As $N$ increases,
the standard deviation (expressed by the error bars) is suppressed and 
it becomes noticeable that $\langle \sigma^N(\omega) \rangle$
has a weak peak structure at $\hbar \omega = 2\gamma_1$.
In the right panel of Fig.~\ref{fig:sigma},
the bunched conductivity $\sigma_{N}(\omega)$ has a strong peak structure at $\hbar \omega = 2\gamma_1$, 
even for a small value of $N$, such as 10.
There are two factors relevant to the appearance of this peak structure; 
firstly, direct interband transitions (between the states with the same $k_{r}$) are optically allowed
and secondary, the density of states is enhanced near the states at $r=1$ and $N+1$ for which 
the bandgap is $\sim 2\gamma_1$.
In Appendix~\ref{app:largeN}, we confirm analytically for a large $N$ that 
direct transitions are responsible for the peak structure
while indirect transitions do not.
The proof is performed for the bunched conductivity.
Although the statement above on the two factors 
does not exactly hold (because of broken selection rule) for the layered conductivity, 
the persistence of the peak structure at $\hbar \omega = 2\gamma_1$ suggests that 
the two factors are approximately valid.

It is also seen that 
the layered conductivity is suppressed (monotonically for a large $N$) 
below the peak by decreasing $\hbar \omega$.
Meanwhile, 
the bunched conductivity $\sigma_{N}(\omega)$ is suppressed
at low photon energies below the peak, 
however, returns to unity at zero energy limit.
The difference between these features suggests that 
a lateral polarization (within a layer) weakens due to the electron hopping between layers,
especially in the $\omega=0$ limit, and that 
an out of plane polarization exists.
The latter makes a net in-plane polarization comparable to the monolayer case of $N=1$, 
when it is projected in the lateral direction.

\subsection{Reflectance}

\begin{figure}[htbp]
 \begin{center}
  \includegraphics[scale=1]{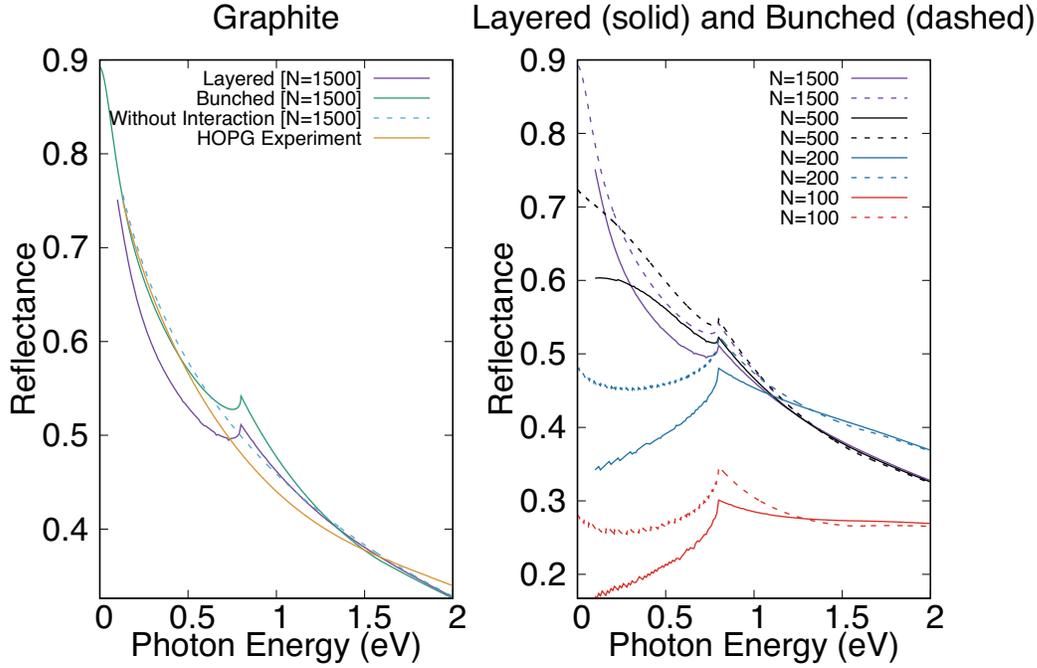}
 \end{center}
 \caption{(Color Online) (left) Comparison between calculated reflectance ($N=1500$) and the experiment on HOPG.
 (right) The peak structure that originates from interlayer interaction 
 appears at $2\gamma_1=0.8$ eV irrespective of the calculation methods. 
 For $\hbar \omega < 2\gamma_1$,
 the layered conductivity underestimates reflectance compared with the bunched conductivity.
 }
 \label{fig:Ref}
\end{figure}

We plot the calculated reflectance as a function of photon energy in Fig.~\ref{fig:Ref}. 
The reflectance was calculated by using the transfer matrix method as explained in Sec.~\ref{sec:tm}.
In the left panel of Fig.~\ref{fig:Ref},
we compare the calculated results for $N=1500$
with the experiments on highly oriented pyrolithic graphite (HOPG).~\cite{Taft1965,Djurisic1999}
We can get some conclusions, 
first, it is difficult to consider that HOPG is dominated by AB stacking, 
because the peak structure does not appear at $2 \gamma_1$ in the measured reflectivity.~\cite{Djurisic1999}
On the other hand, the corresponding peak was observed for natural graphite,~\cite{Taft1965,Hanfland1989a}
and the value of $\gamma_1$ is estimated to be a reasonable value 0.4 eV.
We therefore believe that AB stacking is dominated indeed in natural graphite.
However, the following points need further clarifications on the low-energy structure of the conductivity.
As shown in Fig. 2, 
the layered conductivity decreases monotonically below the peak,
while the bunched conductivity recovers $\pi \alpha$ at the zero energy limit.
The conductivity deduced for natural graphite by Taft and Pilip
(Fig.~6 in Ref.~\citen{Taft1965})
is consistent with this behavior of the layered conductivity. 
However, this consistency should not be used to 
immediately determine the validity of the description of the layered conductivity,
because AA stacking may be present in natural graphite.
As shown in Appendix~\ref{app:aa},
AA stacking increases layered conductivity at low photon energies, while suppresses bunched conductivity.
Detailed analysis when AB and AA stackings are mixed in an $N$-layer graphene deserves a further study.

The results for small $N$ below 1500 are shown in the right panel of Fig.~\ref{fig:Ref}.
The curves obtained from the two conductivities show that 
similar peak structures originating from an interlayer electronic interaction appear
in the reflectance of an $N$-layer graphene at any $N$.
Below the peak, 
the layered conductivity underestimates reflectance compared with the bunched conductivity.

Hanfland {\it et al.} observed that a peak energy position in reflectance of natural graphite
increased with increasing pressure.~\cite{Hanfland1989a}
The observations can be explained as an enhancement of $\gamma_1$, 
due to the pressure induced contraction of $d$.
They also observed that the peak is split into two peaks (denoted by $A_1$ and $A_2$).
The splitting was also observed by means of thermoreflectance measurements at atmospheric pressure.~\cite{Bellodi1975}
We attribute the splitting to a bond alternation along the $c$-axis.
In other words, when interlayer distance $d$ is not exactly uniform but is locally modulated
by a certain lattice distortion like a polyacetylene,~\cite{Su1980,Heeger1988}
an asymmetry between the magnitudes of the two extremal mass
($m_1$ and $m_N$) may arise. 
For example, a correction to $\gamma_1$ of the form
$\delta \gamma_1 \cos(k_r)$
or $m_r \to \gamma_1 \cos(k_r) + \delta \gamma_1 \cos^2(k_r)$ in Eq.~(\ref{eq:bunchedc})
can be used for phenomenological explanations of the double peaks.
An asymmetry between $m_1$ and $m_N$ is also indicative of a broken particle-hole symmetry 
of the band structure.

\subsection{Nonlocal correction}

The underestimation of reflectance 
seen for the layered conductivity in Fig.~\ref{fig:Ref} at low photon energies below the peak
may be related to a nonlocal effect. 
We define the nonlocal conductivity $\sigma^N_{jj'}$ as the corrections to the current of $j$th layer 
that are caused by the electric fields of the other $j'$th layers ($j' \ne j$) as
\begin{align}
 J_j = \sigma^N_{j} E_{j}+ \sum_{j'\ne j} \sigma^N_{jj'} E_{j'}.
\end{align}
The existence of a nonlocal conductivity is physically plausible because 
an electron-hole pair locally excited by an electric field at a layer 
may recombine at a different layer.
Since such a carrier movement is caused by $\gamma_1$, 
the nonlocal effect cannot be significant when $\hbar \omega \gg 2\gamma_1$.

The formulation for the transfer matrix theory based on layered conductivity 
is adaptable to the general case of a nonlocal response.
The boundary condition is generalized from Eq.~(\ref{eq:bc}) to
\begin{align}
 \begin{pmatrix}
  E'_1 \cr
  B'_1 \cr
  E'_2 \cr
  B'_2 \cr
  \vdots \cr
  E'_N \cr
  B'_N
 \end{pmatrix}=
 \begin{pmatrix}
  1 & 0 & 0 & 0 & \cdots & 0 & 0 \cr
  -\frac{\sigma_{1}^N}{\epsilon_0c^2} & 1 & - \frac{\sigma^N_{12}}{\epsilon_0c^2} & 0 & \cdots & - \frac{\sigma^N_{1N}}{\epsilon_0c^2} & 0 \cr
  0 & 0 & 1 & 0 & \cdots & 0 & 0 \cr
  -\frac{\sigma^N_{21}}{\epsilon_0c^2} & 0 & -\frac{\sigma_{2}^N}{\epsilon_0c^2} & 1 & \cdots & - \frac{\sigma^N_{2N}}{\epsilon_0c^2} & 0 \cr
  \vdots & \vdots & \vdots & \vdots & \ddots & 0 & 0 \cr
  0 & 0 & 0 & 0 & 0 & 1 & 0 \cr
  -\frac{\sigma^N_{N1}}{\epsilon_0c^2} & 0 & -\frac{\sigma^N_{N2}}{\epsilon_0c^2} & 0 & \cdots & -\frac{\sigma_N^N}{\epsilon_0c^2} & 1 
 \end{pmatrix}
 \begin{pmatrix}
  E_1 \cr
  B_1 \cr
  E_2 \cr
  B_2 \cr
  \vdots \cr
  E_{N} \cr
  B_{N}
 \end{pmatrix}.
 \label{eq:Apart}
\end{align}
This must be solved together with Eq.~(\ref{eq:To}) or 
\begin{align}
 \begin{pmatrix}
  E_2 \cr
  B_2 \cr
  \vdots \cr
  E_{N} \cr
  B_{N}
 \end{pmatrix}
 =
 \begin{pmatrix}
  \cos(\frac{\omega d}{c}) & i c \sin(\frac{\omega d}{c}) & \cdots & 0 & 0 \cr
  \frac{i}{c} \sin(\frac{\omega d}{c})  & \cos(\frac{\omega d}{c}) & \cdots & 0 & 0 \cr
  \vdots & \vdots & \ddots & 0 & 0 \cr
  0 & 0 & 0 & \cos(\frac{\omega d}{c}) & i c \sin(\frac{\omega d}{c}) \cr
  0 & 0 & 0 & \frac{i}{c} \sin(\frac{\omega d}{c})  & \cos(\frac{\omega d}{c})
 \end{pmatrix}
 \begin{pmatrix}
  E'_1 \cr
  B'_1 \cr
  \vdots \cr
  E'_{N-1} \cr
  B'_{N-1}
 \end{pmatrix}.
 \label{eq:Bpart}
\end{align}
By eliminating $(E'_1,B'_1,\cdots,E'_{N-1},B'_{N-1})^t$
from Eqs.~(\ref{eq:Apart}) and (\ref{eq:Bpart}),
we obtain a self-consistent equation of $(E_2,B_2,\cdots,E_{N},B_{N})^t$.
Numerical calculations tell us $(E'_N,B'_N)^t$ 
for $(E_1,B_1)^t=(1,0)$ or $(0,1)$.
Therefore, we can find a $2\times 2$ matrix $M$ that satisfies 
\begin{align}
 \begin{pmatrix}
  E'_{N} \cr B'_{N}
 \end{pmatrix}
 = M
 \begin{pmatrix}
  E_{1} \cr B_{1}
 \end{pmatrix},
\end{align}
where $M$ includes the nonlocal correction to Eq.~(\ref{eq:ebmatrix}).

\begin{figure}[htbp]
 \begin{center}
  \includegraphics[scale=0.8]{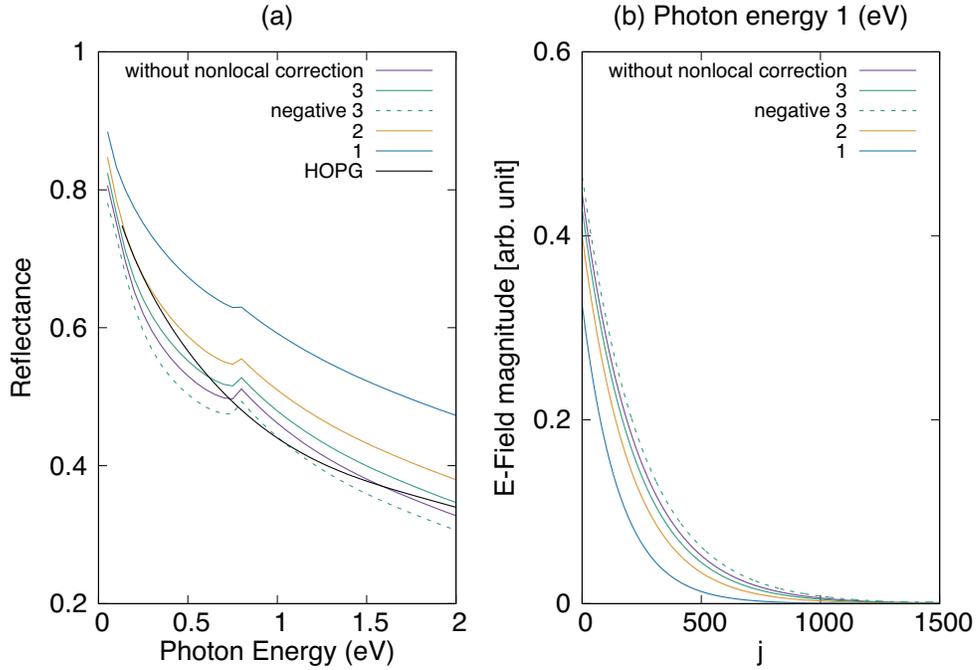}
 \end{center}
 \caption{(Color Online) The effect of nonlocal conductivity on reflectance (a) and electric field (b).
 A positive nonlocal conductivity increases reflectance (a) because it decreases field strength (b).
 The number indicates the $\beta$ value in $\sigma_{jj'}=\pi \alpha \epsilon_0 c e^{-\beta|j-j'|}$, 
 and ``negative'' means that $\sigma_{jj'}=-\pi \alpha \epsilon_0 c e^{-\beta|j-j'|}$.
 A negative nonlocal conductivity decreases reflectance and increases field strength, which is in contrast to the positive case.
 }
 \label{fig:nonlocal}
\end{figure}

Figure~\ref{fig:nonlocal}(a) shows the nonlocal effect on reflectance with $N=1500$.
We obtained the results by employing a simple model 
$\sigma^N_{jj'}=\pi \alpha \epsilon_0 c e^{-\beta|j-j'|}$, 
where a large $\beta$ suppresses the nonlocal effect.
Although the validity of this model is questionable
(because we are not able to calculate it from first principle),
the results help us to understand the way 
in which the observables are changed by nonlocal effects.
The positive (negative) nonlocal conductivity increases (decreases) reflectance, 
and decreases (increases) the electric field as shown in Fig.~\ref{fig:nonlocal}(b).
We consider that a positive nonlocal conductivity is more reasonable 
since the bunched conductivity is larger than the layered conductivity.
For example, when the calculated $E_j$ is sufficiently uniform in all layers, 
we anticipate that the following approximate relationship between the bunched and layered conductivities holds,
\begin{align}
 \sigma_N d \approx \langle \sigma^N \rangle + \frac{1}{N} \sum_j \sum_{j'\ne j} \sigma^N_{jj'}.
\end{align}
This suggests a positive nonlocal conductivity $\frac{1}{N} \sum_j \sum_{j'\ne j} \sigma^N_{jj'}>0$
for a low photon energy (see Fig.~\ref{fig:sigma}).

The underestimation of reflectance of graphite that we have shown for the layered conductivity in Fig.~\ref{fig:Ref}(left)
can be explained by a positive nonlocal conductivity with $\beta =2 \sim 3$.
The $\beta$ value is also reasonably understood with 
the ratio of a lateral transfer integral ($\simeq$ 3 eV) to $\gamma_1$.

\subsection{Universal layer number}

\begin{figure}[htbp]
 \begin{center}
  \includegraphics[scale=1.0]{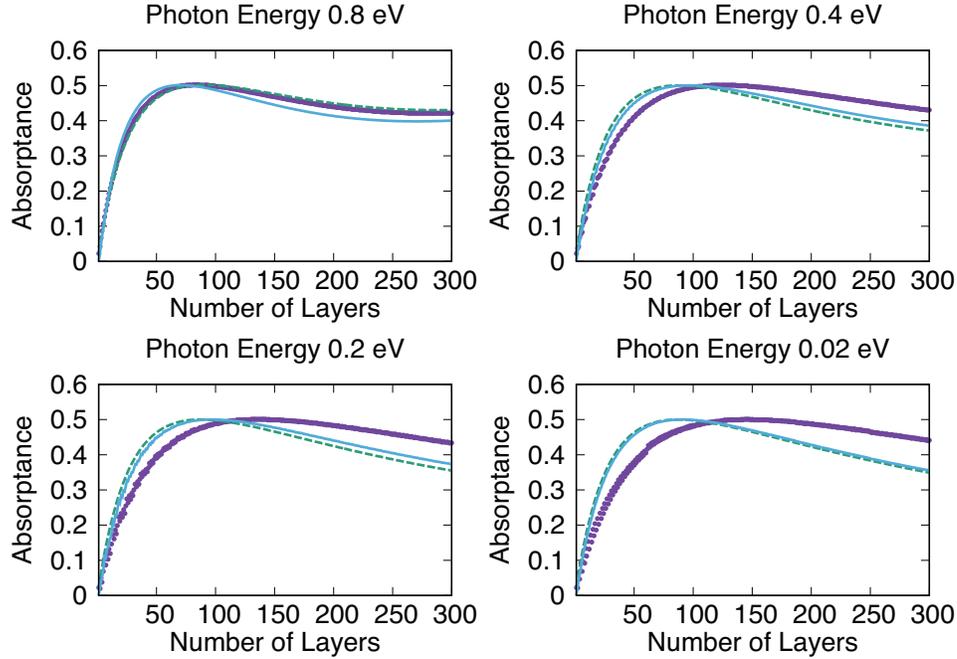}
 \end{center}
 \caption{(Color Online) Absorptance $A^N$ is plotted as a function of number of layers $N$ for several photon energies. 
 The solid/dashed curves are the results calculated with the bunched conductivity.
 These curves have a peak structure at the universal layer number, regardless of the presence or absence of the 
 interlayer interaction.
 Only for the layered conductivity, 
 the interlayer interaction gives a correction to the peak position, 
 and the correction is enhanced in the $\hbar \omega \to 0$ limit.
 }
 \label{fig:absorp}
\end{figure}

We plot the absorptance $A^N$ [Eq.~(\ref{eq:AN})] as a function of $N$ in Fig.~\ref{fig:absorp},
for several photon energies $\hbar \omega =0.8$, $0.4$, $0.2$, and $0.02$ eV.
The dots are obtained by using the layered conductivity, 
while the solid curve is given by the bunched conductivity with Eq.~(\ref{eq:cr0}).
As a reference,
we also show the absorption calculated without an interlayer interaction by the dashed curve which 
has a characteristic peak structure at $N_u \equiv 2/\pi \alpha \simeq 87$.
This is a universal layer number 
because $N_u$ is independent of materials parameters such as Fermi velocity and hopping integral, 
but merely determined by the fine-structure constant $\alpha$ regardless of the frequency.~\cite{Sasaki2020a}
Indeed, when $\gamma_1=0$, by taking $\omega \to 0 $ limit of Eq.~(\ref{eq:ebmatrix}),
we obtain 
\begin{align}
 A^N = \frac{2 N_u N}{(N+N_u)^2},
\end{align}
from which we can readily derive the peak position of the universal layer number.
For the layered conductivity, 
a peak position is shifted by an interlayer interaction and the deviation from $N_u$ increase with decreasing photon energy. 
This is in sharp contrast to the result for the bunched conductivity
for which an interlayer interaction does not significantly shift the peak position and 
the shift is maximum when $\hbar \omega = 2\gamma_1$.~\cite{Sasaki2020a}

\begin{figure}[htbp]
 \begin{center}
  \includegraphics[scale=1.0]{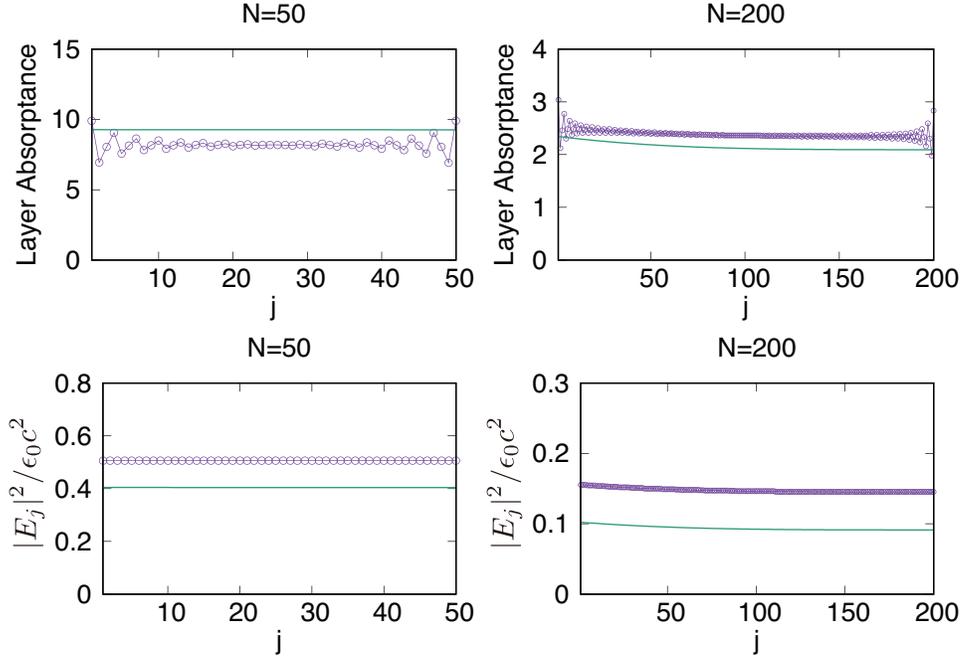}
 \end{center}
 \caption{(Color Online) The spatial dependence of the layer absorptance and field strength 
 are plotted for $N=50$ and 200.
 The interlayer interaction increases the electric field strength, 
 while it suppresses the layered dynamical conductivity (see Fig.~\ref{fig:sigma}).
 The field enhancement is more pronounced for larger $N$.
 The thin curves are the results calculated with $\gamma_1=0$.
 }
 \label{fig:la}
\end{figure}

The mechanism of a change in the peak position may be explained by 
an $N$-dependent enhancement of the electric field, in the following manner.
We take the results of ``Photon Energy 0.4 eV'' in Fig.~\ref{fig:absorp} for explanation, 
where $A^N$ is decreased by the interlayer interaction when $N=50$ and increased when $N=200$.
The layered conductivity when $N=50$ is already converged sufficiently, as we have seen in Fig.~\ref{fig:sigma},
and the average value when $N=200$ is almost the same as the value of $N=50$.
Therefore, by referring to Eq.~(\ref{eq:AN}), we can know that 
the electric field strength must play a key role in explaining their difference.
We plot absorptance by each layer (layer absorptance $A^N_j$) 
and electric field strength ($|E_j|^2$) in Fig.~\ref{fig:la}.
The interlayer interaction always increases the electric field strength.
When $N=50$, the suppressed conductivity overcomes the enhanced electric field,
and the layer absorptance is decreased by interaction.
When $N=200$, the enhanced electric field overcomes the suppressed conductivity,
and the layer absorptance is increased by the interaction.
Thus, graphite may exhibit a fairly complicated depolarization effect 
that depends on $N$ and photon energy.

From Fig.~\ref{fig:la}, we can also find that 
the light absorption in each layer tends to vary spatially, especially near the front and rear layers.
This indicates that the layers near the boundaries are intrinsically more unstable 
than the interior regarding to heating.

\section{Conclusion}\label{sec:con}

We have formulated the layered and bunched conductivities of an $N$-layer graphene in 
Eqs.~(\ref{eq:sigma_j}) and (\ref{eq:bunchedc}), respectively, which are given 
in a simple form as a summation over the ``mass'' variable.
By combining the conductivities with the transfer matrix method, 
we have obtained the optical properties of an $N$-layer graphene.
The calculated reflectance without the interlayer interaction 
is in reasonable agreement with the observed reflectance of HOPG.
An interlayer interaction leads to a peak structure 
in reflectance of an $N$-layer graphene with AB stacking order, for a general $N$.
The peak originates from the direct interband transitions 
for the states near $k=0$ and $\pi$.
Natural graphite exhibits such a peak structure, while HOPG does not.
This indicates the dominance of AB staking in natural graphite, while it is minority in HOPG.
The degeneracy of the two principal direct transitions for the states near $k=0$ and $\pi$
may be resolved by a bond alternation caused by an inhomogeneous interlayer distance.
This will be a key concept in explaining the splitting of 
the peak observed for natural graphite under high pressure. 
For the layered conductivity, we have investigated the effect of nonlocal conductivity on reflectance
and shown that the electromagnetic field is expelled from the graphite by a positive nonlocal conductivity 
so that reflectivity tends to increase. 
This is essential in explaining the discrepancy between theory and measurements.
The behavior of the absorptance $A^N$ near the universal layer number $N_u$ 
is informative in testing the validity of the layered and bunched conductivities.

\begin{acknowledgment}

The author thanks K. Hitachi for developing a numerical program for the transfer matrix method.

\end{acknowledgment}

\appendix

\section{An analytic expression for $\sigma_\infty$}\label{app:largeN}

\begin{figure}[htbp]
 \begin{center}
  \includegraphics[scale=1.0]{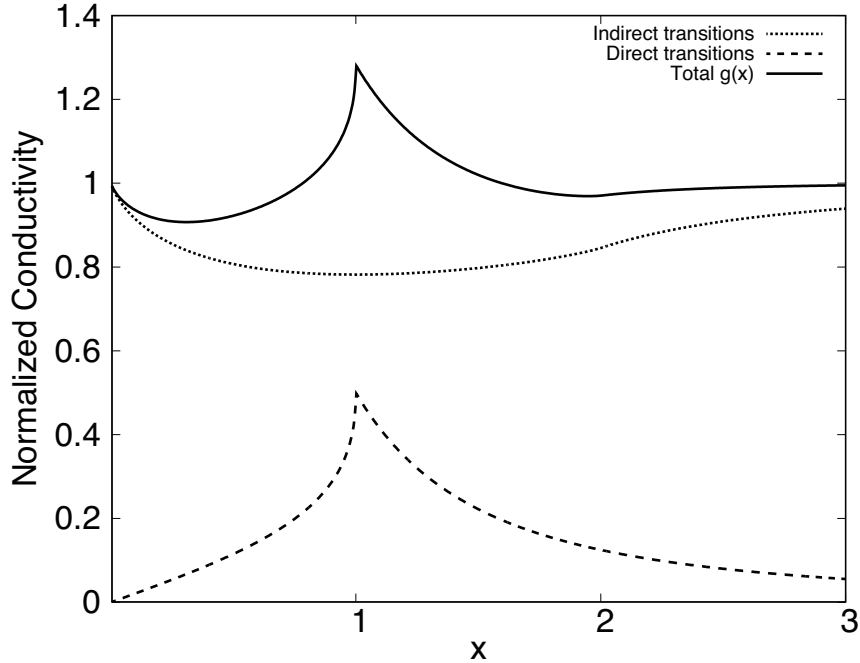}
 \end{center}
 \caption{
 The peak structure (at $x=1$) in the conductivity is attributed 
 to the direct transitions of the states with a high density of states near $k=0$ and $\pi$, 
 for which the transition energy is given by $2\gamma_1$.
 The indirect transitions are the main composition of $g(x)$ for a general $x$.
 The contributions from the indirect transitions are important in the limit $x\to 0$, 
 because the direct transitions are suppressed there.}
 \label{fig:infiniteN}
\end{figure}

We show an analytic expression for the bunched dynamical conductivity $\sigma_N(\omega)$
in the large $N$ limit.
We consider that the result is applicable for $N\ge 50$, because 
$\sigma_{N=50}(\omega)$ is almost converging to $\sigma_{N=1500}$, as shown in Fig.~\ref{fig:sigma} (right).
In this Appendix, 
we use dimensionless variable $x \equiv \frac{\hbar \omega}{2\gamma_1}$ instead of $\omega$.
The bunched conductivity is written as 
\begin{align}
 \sigma_\infty(x) = \frac{\pi \alpha \epsilon_0 c}{d} g(x),
\end{align}
where $g(x)$ is a function defined separately in the following regions.
For $0 \le x \le 1$,
\begin{align}
 & \frac{1}{2x^2} \left\{ 1-\frac{2}{\pi}\cos^{-1}(x)- \frac{2}{\pi}x\sqrt{1-x^2} \right\}  
 \nn \\
 & +1- \frac{2}{\pi} \cos^{-1}\left(\frac{x}{2}\right) - \frac{1}{\pi \sqrt{x^{-2}-1}}
 \left\{
 \ln \left|
 \sqrt{x^{-2}-1}-x^{-1} \right|-
 \ln \left| \frac{\sqrt{x^{-2}-1}\sin(\cos^{-1}(\frac{x}{2})) + \frac{x}{2} - x^{-1} }{\frac{1}{2}} \right| \right\}
 \nn \\
 & + 1 - \frac{2}{\pi \sqrt{x^{-2}-1}} \left\{ \tanh^{-1}\left(\frac{x^{-1}}{\sqrt{x^{-2}-1 }}\right)-
 \tanh^{-1}\left(\frac{x^{-1}+1}{\sqrt{x^{-2}-1 }}\right)
 \right\}.
\end{align}
For $1 \le x$,
\begin{align}
 \frac{1}{2x^2} + 
 1 - \frac{2}{\pi \sqrt{1-x^{-2}}} \left\{ \tan^{-1}\left(\frac{1+x^{-1}}{\sqrt{1-x^{-2} }}\right)-
 \tan^{-1}\left(\frac{x^{-1}}{\sqrt{1-x^{-2} }}\right)
 \right\}.
\end{align}
For $1 \le x \le 2$,
\begin{align}
 1- \frac{2}{\pi} \cos^{-1}\left(\frac{x}{2}\right) 
 - \frac{2}{\pi \sqrt{1-x^{-2}}}
 \left\{  \tan^{-1}\left( \sqrt{\frac{x+1}{x-1}} \right)-\tan^{-1}\left( \sqrt{\frac{x+1}{x-1}}
 \tan\left( \frac{\cos^{-1}\left(\frac{x}{2}\right)}{2} \right)\right) \right\}.
\end{align}
For $2 \le x$,
\begin{align}
 1 - \frac{2}{\pi \sqrt{1-x^{-2}}} \tan^{-1}\left( \sqrt{\frac{x+1}{x-1}} \right).
\end{align}
The peak structure shown in the right panel of Fig.~\ref{fig:sigma} (for $N=1500$)
is given by the following parts in the above expression,
\begin{align}
 g^{\rm direct}(x)
 = \frac{1}{2x^2} \left\{ 1-\frac{2}{\pi}\cos^{-1}(x)- \frac{2}{\pi}x\sqrt{1-x^2} \right\}  \Theta(1-x) + 
 \frac{1}{2x^2} \Theta(x-1).
\end{align}
This originates from the direct transitions $\sigma_{\infty}^{\rm direct}(x)$ in Eq.~(\ref{eq:direct}).
The other remaining parts originate from the indirect transitions $\sigma_{\infty}^{\rm indirect}(x)$.
We show in Fig.~\ref{fig:infiniteN} the total $g(x)$ and the compositions.

\section{AA stacking}\label{app:aa}

Here, we summarize the results of AA stacking. 
While it is known that AA stacking is an unstable structural phase,
we think it is meaningful to investigate the optical properties of AA stacking
in order to see the dependence of the dynamical conductivity on stacking order. 
Moreover, there is a possibility that 
light causes thermal expansion which may drive a transition from AB to AA stacking.

\begin{figure}[htbp]
 \begin{center}
  \includegraphics[scale=1.0]{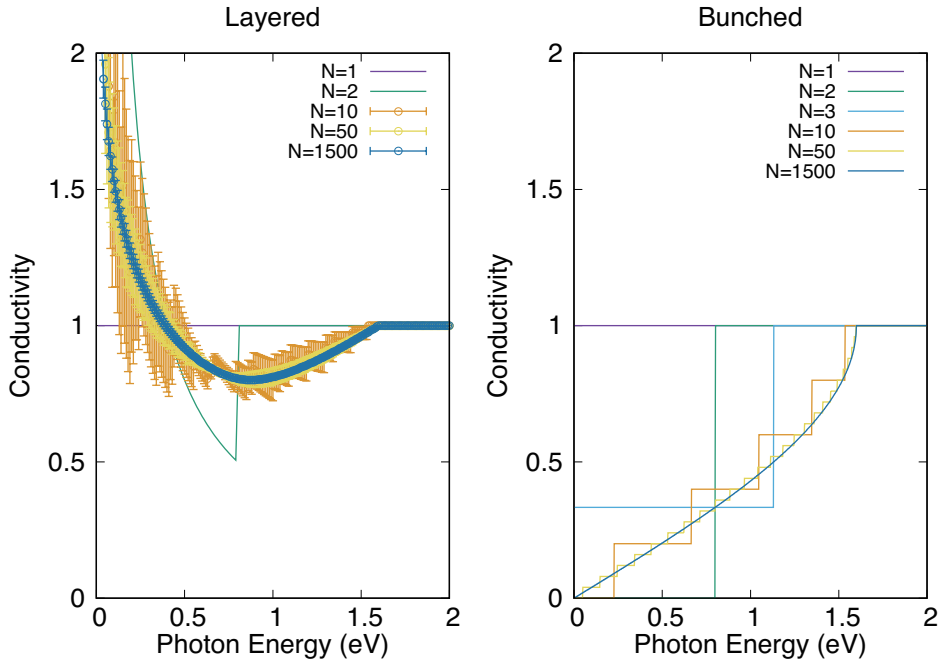}
 \end{center}
 \caption{(Color Online) Layered (left) and bunched (right) conductivities are plotted as a function of
 photon energy $\hbar \omega$ for an $N$-layer graphene with AA stacking.
 The layered and bunched conductivities are scaled by $\sigma^1_1$ or $\sigma_1 d$, respectively.
 The error bars denote the standard deviation.
 }
 \label{fig:sigma_AA}
\end{figure}

We redefine the mass term as two times larger than that of the AB stacking as
\begin{align}
 m_r = 2\gamma_1 \cos (k_r).
\end{align}
The mass appears only as a potential mass in the energy spectrum,~\cite{Min2008}
\begin{align} 
 \varepsilon_{r{\bf p}}^{s} = m_r + s p.
\end{align}
The absence of bandgap means that 
the pseudospin remains intact by AA stacking, which is in sharp contrast to AB stacking.
The pseudospin of monolayer graphene, with the direct product of the standing wave along the $c$-axis, 
constructs the wavefunction.
As a result, for the bunched conductivity, 
the momentum selection rule allows only the momentum preserving ($r'=r$), 
interband ($s'=-s$) transitions.
Therefore, it takes a simple form as
\begin{align}
 \sigma_N(\omega) = \frac{S}{V}g_s g_v
 \frac{e^2}{16\hbar} \sum_{r=1}^{N} \left\{
 \Theta \left( m_r + \frac{\hbar \omega}{2} \right)
 - \Theta \left( m_r - \frac{\hbar \omega}{2} \right)  \right\}.
\end{align}
Whereas, for the layered conductivity,
the momentum selection rule is broken and 
various transitions are allowed as far as energy conservation is satisfied.
Moreover, because the band index $s$ does not separate states into 
the positive (conduction) and negative (valence) energy states,
not only interband transitions ($ss'=-1$) but also intraband transitions ($ss'=+1$) are allowed.
We focus on the real part of the conductivity, for which the contribution from the intraband transitions 
is negligible.
The layered conductivity is given by
\begin{align}
 \sigma^N_j(\omega) &= g_s g_v \frac{e^2}{4\hbar}
 \sum_{r,r'} \left( 1 + \frac{m_{r'}-m_r}{\hbar \omega}\right)
 \frac{\sin^2(jk_r)}{N+1}
 \frac{\sin^2(jk_{r'})}{N+1} \nn \\
 &\times 
 \left\{
 \Theta(m_{r'}+m_r+\hbar \omega) - \Theta(m_{r'}+m_r-\hbar \omega)
 \right\} \nn \\
 &\times
 \left\{
 \Theta(m_{r'}-m_r+\hbar \omega) - \Theta(m_{r}-m_{r'}-\hbar \omega)
 \right\}.
\end{align}
Even though these are categorized into interband transitions, 
many low energy transitions are possible and these 
make layered conductivity having a structure similar to the Drude peak in the $\omega=0$ limit.
We show the calculated conductivities in Fig.~\ref{fig:sigma_AA}.

\bibliographystyle{jpsj}

\end{document}